\documentstyle[prb,aps,epsfig]{revtex}
\def\E{{\rm e}}
\def\I{{\rm i}}
\def\D{{\rm d}}

\begin{document}

\title{Mean-Field Treatment of the Many-Body Fokker-Planck Equation}
\author{Nicolas Martzel\footnote
  {{\bf e-mail:} martzel@gps.jussieu.fr} and Claude Aslangul\footnote
  {{\bf e-mail:} aslangul@gps.jussieu.fr}}
\address{Groupe de Physique des Solides, Laboratoire associ\'e au
CNRS UMR 7588, \\ Universit\'es Paris 7 \&
Paris 6, Tour 23, 2 Place Jussieu, 75251 Paris Cedex 05, France}

\date{\today}

\maketitle
\begin{abstract}

We review some properties of the stationary states of the
Fokker - Planck equation for $N$ interacting particles within a mean
field approximation, which yields a non-linear integrodifferential equation
for
the particle density. Analytical results show that for attractive long range
potentials the steady state is always a precipitate containing one cluster of
small size.
For arbitrary potential, linear stability analysis allows to state the
conditions under
which the uniform equilibrium state is unstable against small perturbations
and,
{\it via} the Einstein relation,
to define a critical temperature $T_{\rm c}$ separating two phases, uniform
and
precipitate. The corresponding phase diagram turns out to be strongly
dependent on the pair-potential.
In addition, numerical calculations reveal that the transition is hysteretic.
We finally discuss the dynamics of relaxation for the uniform state
suddenly cooled
below $T_{\rm c}$.
\end{abstract}


\section{Introduction}\label{intro}

The dynamics of brownian particles is a subject of great interest in
statistical physics, especially when the particles
interact through potentials that can be short or long ranged, from
the simple hard-core to coulombian interactions.
The physics of surfaces, more precisely the motion of adatoms
on substrates, provides an experimental realization of this problem.
Interactions can lead
to collective phenomena, and, generally speaking, to patterns in space.
Pioneering works have revealed experimental and computational
evidences of a phase transition in the case of oxygen adsorbed on
tungsten\cite{Williams,Ching}.
This phase transition and the dynamics of such systems ({\it e. g.} the
modification of the diffusion constant $D$ of a tracer,
time auto-correlation fonction of the on-site density, etc\ldots)
where analyzed by Tringides {\it et al.} \cite{Tring1} and reviewed by
Gomer \cite{Gomer}.
Numerical studies for a two-dimensional
lattice gas have also been performed in the case of a contact interaction
\cite{Tring2,Mag,Gortel}. On a different scale, the hydrodynamics
of interacting brownian
particles have been studied for
short-ranged or screened interactions in\cite{Hess,Nagele}.
In a slighly
different context, especially when the considered system is an open one, many
papers have been devoted to statistical models describing chemical
reactions\cite{asl96} (and references therein, especially the review by Zhdano
and Kasemo\cite{zhdkase}). For details on 2-dimensional lattice gas models
(and
their subsequent approximations), the review by Kehr and Binder \cite{Kehr} is
highly recommendable.

In the present paper, we use the framework of the many-body
Fokker-Planck equation (FPE),  written for $N$ interacting particles -- as
such,
this is a continuous space model.
The ordinary FPE describes the diffusive motion of
brownian
particles under the assumption of {\it slow} diffusion. It can be
derived by using the Kramers-Moyal systematic expansion\cite{gardiner} and
arises
when one assumes that all the moments of the increments of the stochastic
variable of
order $\ge 3$ are proportional to $\Delta t^{r},\,\,r> 1$, where
$\Delta t$ is the time increment. As such, FPE is a conservation
equation for a probability density $P$, which can always be written in the
form:
\begin{equation}
     \partial_{t} P\,=\,-\,\mbox{div }J
     \enspace, \label{FPE}
\end{equation}
where $J$ is the probability current. As an example, for a single particle
with
position ${\bf x}$ moving in the static external potential $V(\bf x)$, the
current is the sum of the drift term $-\mu P\partial_{\bf x} V$ and of the
diffusion current $-D \partial_{\bf x} P$; in such a case, the equation
writes:
\begin{equation}
     \partial_{t}P({\bf x},\,t)\,=\,\partial_{\bf x}
     \left[D\,\partial_{\bf x}\,P({\bf x},\,t)+\mu\, P({\bf
x},\,t)\partial_{\bf x}V({\bf x})\right]
     \enspace, \label{FPE1}
\end{equation}
where $\mu$ is the mobility and $D$ the diffusion constant. Eq. (\ref {FPE1})
gives the probability density for the position of a
particle obeying the Langevin equation in its viscuous limit, an
initial distribution being given.

With several particles, the interesting case occurs when they interact
through a given internal force field deriving from a potential $V$, opening
the
possibility of competing effects. When $V$ is purely repulsive, no
interesting effect
is expected: in
infinite space, one can guess that the equilibrium state is the uniform one,
all probability densities being
constant in space. In the opposite case, when the particles attract
each other, a competition between diffusion and interaction takes
place, which can, at least in principle, produce patterns or
structures in space. Obviously enough, the possibility of the latter
depends on the features of the interaction potential, namely its
strength and its range, and possibly of the dimensionality. The competition
between drift and diffusion is
measured by the ratio $D/\mu$; as a consequence, patterns
can be expected at low temperatures, {\it i. e.} when the drift term
dominates diffusion. If there exists a definite value $D_{\rm c}$ of $D$
which separates two
distinct stationary solutions, the Einstein relation $D/\mu=k_{\rm B}T$
allows to
identify a critical temperature $T_{\rm c}$.

The purpose of this paper is to put forward a
few results concerning the equilibrium state of FPE for interacting
particles, obtained within a mean field approximation. The paper is
organized as
follows. After setting the basic equations relevant to our purpose, we first
focus on some potentials allowing an exact treatment of the
mean-field equations. In a second step, we discuss the linear stability
of the uniform equilibrium state, which is always a solution of the
problem. It is seen that unstabilities can indeed occur, and the
conditions for that are given, yielding the expression of the critical
temperature $T_{\rm c}$. Eventually, we give the far-from-equilibrium dynamics
of the uniform state suddenly cooled below $T_{\rm c}$.

For $N$ identical interacting particles, the potential is noted $V({\bf
x}_{1},\,{\bf
x}_{2},\ldots\,{\bf x}_{N})$ and the generalization of (\ref{FPE1})
writes:
\begin{equation}
     \partial_{t} P({\bf x}_{1},\, \ldots ,\,{\bf x}_{N},\,t)\, =
     \sum_{i} \partial_{{\bf x}_{i}}\left [D \,  \partial_{{\bf x}_{i}}
     P({\bf x}_{1},\, \ldots ,{\bf x}_{N},t) + \mu \,
     P({\bf x}_{1}, \,\ldots ,\,{\bf
     x}_{N},\,t)\,
     \partial_{{\bf x}_{i}} V({\bf x}_{1}, \,\ldots,{\bf x}_{N}) \right]
     \enspace;\label{FPN}
\end{equation}
in the following the potential is assumed to be the sum of $N(N-1)$ even
two-body terms:
\begin{equation}
     V({\bf x}_{1},\,{\bf x}_{2},\ldots\,{\bf x}_{N})\,=\,
     \frac{1}{2}\,\sum_{i\neq j}\,v({\bf x}_{i}- {\bf x}_{j})
     \enspace,\hspace{30pt}v({\bf x})\,=\,v(-{\bf x})
     \enspace. \label{pot}
\end{equation}

Obviously, the solution of (\ref {FPN}) -- if known --, is of little
physical interest, since one is usually interested in the
one-particle density $P^{(1)}$ and the pair-density function $P^{(2)}$ (reduced
densities of
order 1 and 2), defined as:
\begin{equation}
     P^{(1)}({\bf x},\,t)\,=\,\int \D {\bf x}_{2}\ldots\D{\bf x}_{N}
     P({\bf x},\,{\bf x}_{2}, \ldots ,\,{\bf x}_{N},\,t)\enspace,
     \hspace{20pt}
     P^{(2)}({\bf x},\,{\bf x}',\,t)\,=\,\int \D {\bf x}_{3}\ldots\D {\bf
x}_{N}
     P({\bf x},\,{\bf x}',\,{\bf x}_{3},\, \ldots ,\,{\bf
x}_{N},\,t)\enspace.\label{DefRed}
\end{equation}
Due to the many-body interactions,
reduced
densities obey a hierarchy of the BBGKY type (see {\it e. g.}
McQuarrie\cite{macquarrie}); in the present context,
the first equation of this hierarchy writes:
\begin{equation}
     \partial_{t} P^{(1)}({\bf x},\,t) =\,\partial_{{\bf x}}\left[
     D \,\partial_{{\bf x}} P^{(1)}({\bf x},\,t)
     + (N-1)\,\mu  \,
     \int { \D {\bf x}' \,\,\, P^{(2)}
     ({\bf x},{\bf x}',\,t)\,\partial_{{\bf x}}
     v({\bf x} - {\bf x}')} \right] \enspace.\label{FPEred1}
\end{equation}
Solving this hierarchy is usually impossible; the simplest approximation is
of the mean-field type, in which one imposes the form:
\begin{equation}
     P({\bf x}_{1},{\bf x}_{2},\ldots{\bf x}_{N},\,t) \,=\,
     \prod_{i=1}^{N}\,p({\bf x}_{i},\,t)
    \enspace.\label{HypMF}
\end{equation}
From (\ref {FPEred1}), the mean-field one-particle probability density $p$
must obey the following equation:
\begin{equation}
     \partial_{t} p({\bf x},\,t) =\partial_{{\bf x}}\left[
     D \,\partial_{{\bf x}} p({\bf x},\,t)
     + (N-1)\,\mu \, \,p({\bf x},\,t)
    \int { \D {\bf x}' \,p({\bf x}',\,t)\,\partial_{{\bf x}}
     v({\bf x} - {\bf x}')} \right]  \enspace.\label{FPEMF}
\end{equation}
Thus, for $N\gg 1$, the particle density $n({\bf x},\,t)=Np({\bf x},\,t)$
is the
solution of the integrodifferential non-linear equation:
\begin{equation}
     \partial_{t} n({\bf x},\,t) \,=\,\partial_{{\bf x}}\left[
     D \, \partial_{{\bf x}}\,n({\bf x},\,t)
     + \,\mu \, \,n({\bf x},\,t)\,\partial_{{\bf x}}
    \int { \D {\bf x}' \,n({\bf x}',\,t)\,
     v({\bf x} - {\bf x}')} \right]  \enspace.
     \label{FPEMFdens}
\end{equation}
Clearly, even in this mean-field approximation, finding the solution is
by far not a trivial question.

\section{Some exact stationary solutions of the one-dimensional Mean-Field FPE}
\label{exactMF}

In one dimension, all the stationary solutions of
(\ref{FPEMFdens}) give a vanishing current
$J$ and are the solutions of:
\begin{equation}
     D \, \frac{\D }{\D x}\,n(x) + \,\mu \,\,n(x)\,
     \frac{\D }{\D x}\int_{-\infty}^{+\infty} { \D x' \,n(x')\,v(x - x')}\,=\,0
     \enspace.\label{FPEMFJ0}
\end{equation}
Because $n$ also appears in the integral, the stationary mean-field solution
is not connected in an obvious way to the Boltzmann
distribution built with $v(x)$. One could naively believe that,
since in the mean-field treatment each particle interacts with
$N-1\simeq N$
other particles through the potential $v(x)$, its equilibrium
distribution is $\propto \E^{-N\beta v(x)}$. This turns out to be
wrong in general, except for the harmonic potential. Indeed, the
one-particle density is obtained by integrating over the coordinates
of all the other $N-1$ particles and there is no reason ensuring that
the bare two-body interaction $v(x)$ should spontaneously appear
in the Boltzmann way in the one-particle density, even in a mean-field
approach. It will be seen that the ordinary Boltzmann
factor is recovered only in certain limits (see below). Also note
that, since there is no external force field, the equilibrium
states --
assumed to be independent of the initial condition which naturally implies
privileged points  -- are defined up
to an arbitrary translation in space. Otherwise stated, if $n(x)$ is a
solution of (\ref{FPEMFJ0}) defined for all $x$ between $\pm\infty$,
then $n_{x_{0}}(x)=n(x-x_{0})$, $x_{0}$ arbitrary, is also a solution. This
degeneracy is discarded by the use of symmetric boundary conditions (see
below) or by having in mind that the displayed equilibrium states
arise from the initial condition $p(x,\,t=0)=\delta(x)$.

For some potentials, the solution of
the equation
(\ref{FPEMFJ0}) can be found in closed form. In the following section, we
give a few examples and briefly analyze the corresponding solutions.

\subsection{The Coulomb Potential}\label{coulomb}
We first choose:
\begin{equation}
     v(x)\,=\,v_{0}\,\frac{ |x|}{\xi}\enspace,\hspace{30pt}(v_{0}>0)
     \enspace,\label{coul1d}
\end{equation}
which, for $d=1$, mimics
the Coulomb potential in the sense that $v(x)$ satisfies the Poisson
equation. This is clearly a long-ranged attractive potential: the
force exerted on a given particle is constant in space and is equal
to $-(v_{0}/\xi)\,{\rm sgn}x$.

Let us first assume that the particles are confined in the
interval $[-L/2,\,+L/2]$. Introducing the integrated density $ q(x)$:
\begin{equation}
     q(x)\,=\,\int_{0}^{x}  \,n(x')\,\,\D x'
     \enspace,\label{densint}
\end{equation}
it is readily seen from (\ref{FPEMFJ0}) and (\ref{coul1d}) that $q(x)$
obeys the following differential equation ($\beta^{-1}=k_{\rm B}T$):
\begin{equation}
     q''(x)+\frac{\beta v_{0}}{\xi}\,q'(x)\,[2q(x)-N]\,=\,0
     \enspace.\label{eqdifq}
\end{equation}
Using the boundary conditions $q(-L/2)=0$ and $q(+L/2)=N$, a little
algebra yields the properly normalized density in the limit
$L\rightarrow+\infty$:
\begin{equation}
     n(x)\,=\,\frac{N}{2\Delta\cosh^{2}(x/\Delta)}\enspace,\hspace{20pt}
     \Delta\,=\,\frac{2\xi}{N \beta v_{0}}
     \enspace.
     \label{pcoul1d}
\end{equation}
This is a peaked distribution, which tends towards $N \delta(x)$
in the limit $N\rightarrow +\infty$. Due to the infinite
range of the potential, the stationary state is for any $D$ and
large $N$ a cluster
of very small shape as compared to $\xi$ when $\beta v_{0}\sim 1$. Note
that $p(x)$ is flat at $x=0$ but decreases approximately
like an exponential:
\begin{equation}
     n(x)\,\simeq\,\xi^{-1}\,N^{2}\beta v_{0}\,\E^{-N\beta
v_{0} |x|/\xi}\,
     \equiv\,\xi^{-1}\,N^{2}\beta v_{0}\,\E^{-N\beta
     v(x)}\hspace{5pt}\,\mbox{if}\hspace{5pt}\,x\,\gg\,\frac{\xi}{N\beta
v_{0}}\enspace.
     \label{pcoul1dap}
\end{equation}
This shows that the Boltzmann behaviour is recovered in the wings of the
distribution only.

Note that it is not necessary to introduce a finite interval of length
$L$ and subsequently to take the limit $L\rightarrow\infty$. Yet, this
procedure
allows to discuss the invariance by translation mentionned above.
Indeed, taking the boundaries at $\pm
L/2$ forces the solution $n(x)$ to be even, a symmetry which is
conserved when the limit $L\rightarrow\infty$ is performed afterwards.
On the other hand, by assuming infinite space at the beginning, the
calculation yields the same solution (\ref {pcoul1d}) as well as all the
translated functions $n_{x_{0}}(x)=n(x-x_{0})$ with $x_{0}$ arbitrary. It
can be
checked that all the functions $n_{x_{0}}(x)$ indeed satisfy
(\ref{FPEMFJ0}).

\subsection{The Harmonic Potential}\label{harmonic}
We now take $v(x)=v_{0}x^{2}/(2\xi^{2})$, $v_{0}>0$, another
example of
long-ranged attractive potential. In this case, (\ref{FPEMFJ0})
writes:
\begin{equation}
    n'(x) + \,\frac{\beta v_{0}}{\xi^{2}}\,\,n(x)\,
     \int_{-\infty}^{+\infty} \, (x - x')\,\,n(x')\, \D x'\,=\,0
     \enspace.\label{FPEMFharm}
\end{equation}
For even $n(x)=n(-x)$, this simplifies to:
\begin{equation}
    n'(x) + \,N\frac{\beta v_{0}}{\xi^{2}}\,x\,n(x)\,\,=\,0
     \enspace,\label{FPEMFharm1}
\end{equation}
which gives the solution for the density:
\begin{equation}
    n(x) \,=\,\frac{N}{\sqrt{2 \pi} \Delta}
    \E^{-\,x^{2}/(2\Delta^{2})}\enspace,
    \hspace{20pt}\Delta=\frac{\xi}{\sqrt{N \beta v_{0}}}
     \enspace.\label{pharm}
\end{equation}
Note that the expression (\ref{pharm}) is simply of the form $\propto
\E^{-N\beta\,v(x)}$, which is the Boltzmann distribution for a single
particle elastically bound with $N$ others. The fact that this is true for
all $N$
and $x$ is clearly characteristic of the harmonic potential. Again, it is
readily checked that all the
functions $n(x-x_{0})$ satisfy (\ref{FPEMFharm}) with $x_{0}$ arbitrary.

\subsection{Polynomial Potentials}\label{polypot}
The harmonic potential treated above immediately gives the clue for
solving the same problem with any potential of the polynomial
type:
\begin{equation}
    v(x) \,=\,\sum_{r\ge 1}\,c_{r}x^{r}
     \enspace.\label{potpoly}
\end{equation}
With this kind of potential, (\ref{FPEMFJ0}) assumes the form:
\begin{equation}
    n'(x) + \,\beta \,\,n(x)\,
     \sum_{r\ge 1}\,rc_{r}\int _{-\infty}^{+\infty}\, (x -
x')^{r-1}\,\,n(x')\,\D x'\,=\,0
     \enspace.\label{FPEMFpopol}
\end{equation}
By rearranging terms in the integral, this can be rewritten as:
\begin{equation}
    n'(x) + \,\beta \,\,n(x)\,
     \sum_{r\ge 1}\,\gamma_{r}(x)\,f_{r}\,=\,0
     \enspace,\label{FPEMFpopol1}
\end{equation}
where the $\gamma_{r}(x)$ are definite polynomials and where the
quantities $f_{r}$ are the moments of $n(x)$:
\begin{equation}
    f_{r}\,=\,\int_{-\infty}^{+\infty}\,  x^{r}\,n(x)\,\D x
     \enspace.\label{momn}
\end{equation}
Now, (\ref{FPEMFpopol1}) can be formally integrated, to give a function
$p(x)$ containing the parameters $f_{r}$:
\begin{equation}
    n(x) \,=\,\E^{-\beta\,\sum_{r\ge 0}\,\Gamma_{r}(x)\,f_{r}}
     \enspace,\label{FPEMFpopsol}
\end{equation}
the $\Gamma_{r}$ being definite functions depending on $v(x)$. By reporting
this expression
in  (\ref{momn}), one can write as many equations as necessary to
find the $f_{r}$ and eventually obtain the explicit expression of $n(x)$.
Clearly, the latter is not {\it a priori} of the form $\propto
\E^{-N\beta\,v(x)}$.

As an example, let us consider the quartic potential:
\begin{equation}
    v(x) \,=\,v_{0}\,\left[g\left(\frac{x}{\xi}\right)^{2}+
    \left(\frac{x}{\xi}\right)^{4}\right]
     \enspace.\label{potquart}
\end{equation}
This potential is purely  attractive if $g\,>\,0$; otherwise, it is
repulsive for $x$ between $\pm \sqrt{[-g/2]}$ and
attractive elsewhere. Inserting this potential in (\ref{FPEMFpopol})
and integrating, one finds:
\begin{equation}
    n(x) \,=\,C\,\E^{-\beta v_{0}\,[(g\xi^{2}f_{0}+6f_{2})\,x^{2}/\xi^{4}
    +f_{0}\,x^{4}/\xi^{4}]}
     \enspace.\label{FPEMFquart}
\end{equation}
$f_{0}$ is equal to $N$, whereas the unknown quantities $C$ and
$f_{2}$ can be derived from the two equations:
\begin{equation}
    2C\,\int_{0}^{+\infty}\,
    \E^{-\beta v_{0}\,[(gN+f_{2}/\xi^{2})X^{2}+NX^{4}]}\,\D X\,=\,N
     \enspace,\hspace{20pt}
     2C\,\xi^{2}\,\int_{0}^{+\infty}\,X^{2}\,
    \E^{-\beta v_{0}\,[(gN+f_{2}/\xi^{2})X^{2}+NX^{4}]}\,\D X\,=\,f_{2}
     \enspace.\label{FPEMFquartAf2}
\end{equation}
Note that the expression (\ref{FPEMFquart}) is not $\propto
\E^{-N\beta v(x)}$, except if $f_{2}$ is negligible as compared to
$N\xi^{2}$ (see below). The above integrals can be expressed
\cite{Erdelyi} with the Weber functions ${\cal
D}_{\nu}$:
\begin{equation}
    \sqrt{\pi}\,C\,(2N\beta v_{0})^{-1/4}\,\E^{A^{2}/4}\,{\cal
    D}_{-1/2}(A)\,=\,N
     \enspace,\hspace{20pt}
     \frac{\sqrt{\pi}}{2}\,C\,\xi^{2}\,(2N\beta
v_{0})^{-3/4}\,\E^{A^{2}/4}\,{\cal
    D}_{-3/2}(A)\,=\,f_{2}
     \enspace,\label{quartAf2}
\end{equation}
where the constant $A$ is:
\begin{equation}
    A\,=\,\frac{\beta v_{0}}{2N}\,(gN+f_{2}/\xi^{2})^{2}
     \enspace.\label{defAf}
\end{equation}
For arbitrary $N$, these equations can be numerically solved to
provide the two parameters $C$ and $f_{2}$. On the other hand, when
$N$ is very large, one can find directly asymptotic
expressions of the integrals appearing in (\ref{FPEMFquartAf2}). By
doing so and coming back to the mean-field density $n(x)$, one
gets
according to the sign of $g$:
\begin{equation}
    n(x)\,\simeq\,N \xi^{-1}\,\left(\frac{2g\beta
v_{0}N}{\sqrt{\pi}}\right)^{1/2}\,
    \E^{-N\beta v(x)}\hspace{20pt} (g>0)
     \enspace,\label{quartpxgp}
\end{equation}
\begin{equation}
    n(x)\,\simeq\,N \xi^{-1}\,\left(\frac{|g|\beta
v_{0}N}{\sqrt{2\pi}}\right)^{1/2}\,
    \E^{-N\beta v_{0}g^{2}/4}\,\E^{-N\beta v(x)}\hspace{20pt} (g<0)
     \enspace.\label{quartpxgn}
\end{equation}
This shows that the Boltzmann distribution is recovered only in the limit
$N\gg 1$ .
For positive $g$, one has a single cluster
with a width $\sim(N\beta v_{0})^{-1/2}\,\xi$. For negative $g$, two
peaks arise, both having the latter width, and separated by
$\sqrt{-2g}\,\xi$.

Note that, in any case with $\beta v_{0}\sim 1$, the mean-field stationary
state for large $N$ is a compact
cluster with a very small size as compared to the characteristic
length $\xi$ of the potential.

\section{Linear stability analysis of the uniform state}\label{stablan}

In the previous section, we displayed several potentials allowing an
exact explicit expression for the non-uniform equilibrium state with a
vanishing
current. On the other hand, (\ref{FPEMFJ0}) always trivially has the
uniform state as solution. All this means that several solutions can exist and
the question arises to settle which of them can indeed be realized for a given
ratio $D/\mu$. The aim of this section is to analyze the linear
stability of the
uniform solution $n_{0}$ of the
mean-field equation (\ref{FPEMFJ0}) in any dimension. Setting $n({\bf
x},\,t)=n_{0}+\delta n({\bf x},\,t)$ and discarding all terms of order
greater than one, one obtains:
\begin{equation}
     \partial_{t}\delta n({\bf x},\,t)\,=\, D \, \partial_{{\bf
x}}^{2}\,\delta n({\bf x},\,t)
     + \,\mu \, \,n_{0}\,\partial_{{\bf x}}\,
    \int { \D {\bf x}' \,\delta n({\bf x}',\,t)\,
    \partial_{{\bf x}} \,v({\bf x} - {\bf x}')}
    \enspace.\label{FPElin}
\end{equation}
This linear equation can now be analyzed by introducing the Fourier
transforms:
\begin{equation}
     \rho({\bf q},\,t)\,=\,
    \int \, \D {\bf x} \,\E^{-\I {\bf q}{\bf x}}\,\delta n({\bf
    x},\,t)\enspace,\hspace{30pt}
    {\tilde v}({\bf q})\,=\,
    \int \, \D {\bf x} \,\E^{-\I {\bf q}{\bf x}}\,v({\bf x})
    \enspace,\label{Fourier}
\end{equation}
and it is readily seen that all the eigenmodes of (\ref{FPElin}) are of the
form
$\E^{-\omega({\bf q})t}$, with $\omega({\bf q})$ given by the dispersion
relation:
\begin{equation}
     \omega({\bf q})\,=\,\left[ D + \mu\,n_{0}\,{\tilde v}({\bf q})
    \right]\,{\bf q}^{2}\,\equiv\,D_{\rm eff}({\bf q})\,{\bf q}^{2}
    \enspace.\label{disp}
\end{equation}
$D_{\rm eff}$ plays the role of an effective ${\bf q}$-dependent diffusion
constant;
it can be said that, due to interactions,  Fick's law is no more a local
law. Setting:
\begin{equation}
     v(x)\,=\,v_{0}\,\phi(x/\xi)\enspace,\hspace{20pt}
     \phi(0) \,>\,0\enspace,\hspace{20pt}x\,=\,|{\bf x}|
     \enspace,\hspace{20pt}q\,=\,|{\bf q}|
    \enspace,\label{potadim}
\end{equation}
the dispersion law (\ref{disp}) in $d$ dimensions can be rewritten as:
\begin{equation}
     \omega(q)\,=\,D\,\left[ 1 + \beta v_{0}\,n_{0}\,\xi^{d}\,U_{d}(q\xi)
    \right]\,q^{2}
    \enspace,\label{dispnew}
\end{equation}
where $U_{d}$ is a dimensionless function. More precisely, one has ($J_{0}$ is
the ordinary Bessel function):
\begin{equation}
     U_{1}(\kappa)\,=\,2\,\int_{0}^{+\infty}\,\phi(X)\,\cos(\kappa X)\,\D X
    \enspace,\hspace{10pt}
    U_{2}(\kappa)\,=\,2\pi\,\int_{0}^{+\infty}\,X\,\phi(X)\,J_{0}(\kappa
    X)\,\D X
    \enspace,\label{U12}
\end{equation}
\begin{equation}
U_{3}(\kappa)\,=\,\frac{4\pi}{\kappa}\,\int_{0}^{+\infty}\,X\,\phi(X)\,
\sin(\kappa X)\,\D X
    \enspace.\label{U3}
\end{equation}
Eq. (\ref {dispnew}) shows that the uniform solution is
unstable against a
deformation with a wavevector ${\bf q}$ satisfying:
\begin{equation}
     k_{\rm B}T+n_{0}\,\xi^{d}\,v_{0}\,U_{d}(q\xi)\,<\,0
    \enspace.\label{instable}
\end{equation}
Clearly, such a condition cannot be satisfied for a purely repulsive
potential, since then both $v_{0}$ and $U_{d}$ are positive
quantities\footnote{When $\phi(X)$ is a positive function, $U_{d}$ is bounded
below by a positive (possibly diverging) integral.}: for repulsive forces,
the uniform state is always stable, as
anticipated on physical grounds in section \ref{intro}.

When $v$ is attractive at short distances, $v_{0}$ is
negative. The possibility of instability then depends only on the
precise behaviour of the dimensionless Fourier transform $U_{d}$. For
very short ranged potentials, $U_{d}(\kappa)$ is expected to be positive for
all $\kappa$. The instability condition can then be rewritten in a more
transparent form :
\begin{equation}
     k_{\rm B}T\,<\,n_{0}\,\xi^{d}\,|v_{0}|\,U_{d}(q\xi)\enspace,
     \hspace{10pt}v_{0}<0
    \enspace.\label{TCe}
\end{equation}
If, in addition, $U_{d}$ is a monotonous decreasing function,
instability occurs at large wavelengths provided the temperature is
low enough. More precisely, provided that $U_{d}(0)$ is finite
(short-range potential), one can define a critical temperature
$T_{\rm c}$:
\begin{equation}
     k_{\rm B}T_{\rm c}\,=\,n_{0}\,\xi^{d}\,|v_{0}|\,U_{d}(0)
    \enspace.\label{TC}
\end{equation}
For $T\,<\,T_{\rm c}$, there exists a finite interval $[0,\,q_{\rm max}]$
in which the uniform state in unstable. $q_{\rm max}$ is a function
of the density $n_{0}$ and the temperature. Assuming that $U_{d}(0)$
is of order unity, (\ref {TC}) means that the critical temperature is
such that the thermal energy is of the order of the effective
interaction energy.

The existence of an upper $q_{\rm max}$, below which the uniform solution is
unstable is easily understood on physical grounds. For a small disturbance
having
a large wavelength compared to the range of the potential, particles in excess
tend to attract each other more firmly, increasing holes in the density. On
the
other hand, when the wavelength is quite small, the inhomogenities are easily
washed out since holes and particles all interact.

\begin{figure}[htbp]
\centerline{\epsfxsize=9cm\epsfbox{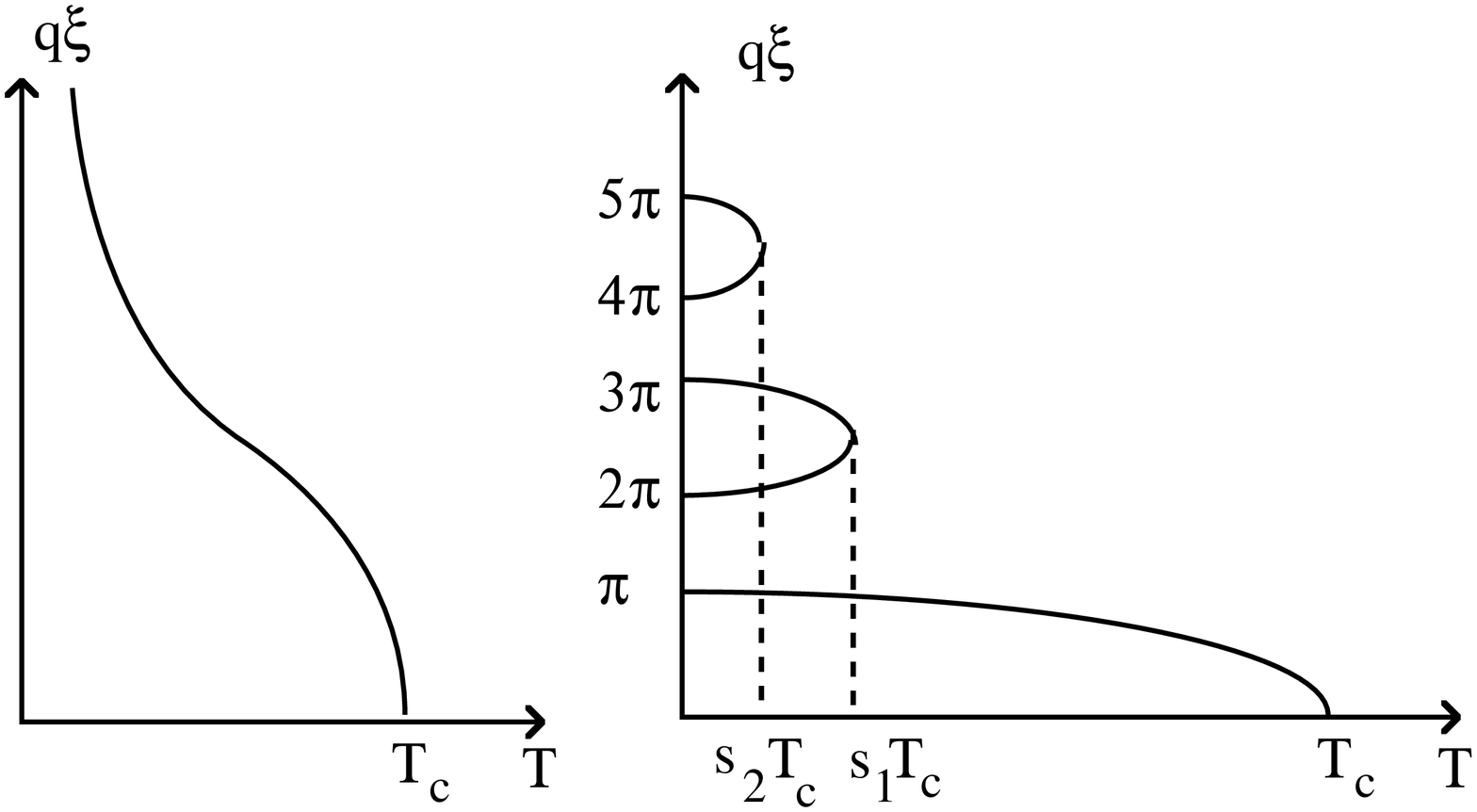}}
\vspace{-40pt}\caption{Schematic representation of the instability domains
for the
gaussian potential (\ref {gaussd}) (left, below $q_{\rm max}$) and for the
square potential (\ref{creneau}) (right, pockets).}
\label{qinstcren}
\end{figure}

As an example, let us consider the $d$-dimensional
gaussian attractive potential:
\begin{equation}
     v(x)\,=\,v_{0}\,\E^{-x^{2}/(2\xi^{2})}\enspace,\hspace{20pt}
     v_{0}\,<\,0
    \enspace.\label{gaussd}
\end{equation}
In this case, one has:
\begin{equation}
     k_{\rm B}T_{\rm c}\,=\,(2\pi)^{d/2}n_{0}\,\xi^{d}\,|v_{0}|
    \enspace\label{Tcgaussd}
\end{equation}
and $q_{\rm max}$ is
given by:
\begin{equation}
     q_{\rm max}\,=\,\xi^{-1}\,\left(2\,\ln \frac{T_{\rm
     c}}{T}\right)^{1/2}\enspace,\hspace{20pt}T\,\le\, T_{\rm c}
    \label{qmaxgaussd}
\end{equation}
which entails that:
\begin{equation}
     q_{\rm max}\,\sim\,(T_{\rm
c}-T)^{1/2}\enspace,\hspace{20pt}T\,\lesssim T_{\rm c}
    \enspace.\label{expoqmaxgd}
\end{equation}
These last results are depicted in the left part of fig. \ref{qinstcren}.
The same conclusions still qualitatively hold if a repulsive core is
added ({\it e. g.} 6-12 Lennard-Jones or Morse potential). This only
affects the large-$q$ behaviour of $\tilde v$ and does not change the
results.

When the potential has a sharp cut-off, its Fourier transform
has an oscillatory behaviour at relatively small wave numbers and an
interesting phenomenon occurs. To be specific, let us take the following
square
potential:
\begin{equation}
     v(x)\,=\,\left\{ \begin{array}{ll}
              v_{0} & \mbox{if $|x|\,<\,\xi/2$}\\
                 0 &\mbox{otherwise}
                       \end{array}\enspace,
               \right.
    \hspace{20pt}v_{0}\,<\,0\enspace.\label{creneau}
\end{equation}
We here find $k_{\rm B}T_{\rm c} =
n_{0}\xi|v_{0}|$. Now, let us note $X_{k}$ the {\it absciss\ae}
\hspace{3pt}of the maxima of $(\sin
X)/ X$ and $s_{k}\equiv (\sin X_{k})/X_{k}$ ($X_{0}=0$, $s_{0}=1$).
Then, for $s_{1}T_{\rm c}\,<\,T\,<T_{\rm c}$, a
single instability
interval arises, $[q_{\rm min,\,1}=0,\,q_{\rm max,\,1}]$. When $T$
decreases, another interval
$[q_{\rm min,\,2},\,q_{\rm max,\,2}]$ is found when $s_{2}T_{\rm
c}\,<\,T\,<\,s_{1}T_{\rm
c}$, and so on. The situation is depicted on fig. \ref{qinstcren}.
Thus, for a short-range potential with a sharp cut-off, several
disjoint instability intervals are successively obtained when the
temperature is decreased.

\begin{figure}[htbp]
\centerline{\epsfxsize=8cm\epsfbox{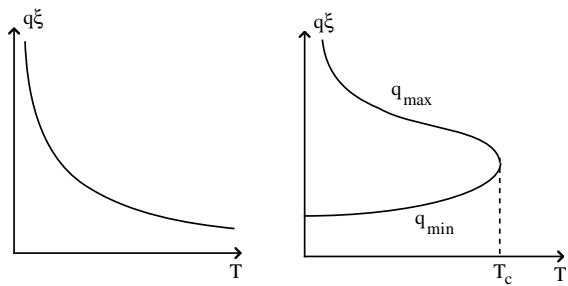}}
\vspace{-40pt}\caption{Schematic representation of the instability domain
for the
Coulomb potential (\ref{coul1d}) (left, below $q_{\rm max}$) and the
inverted Morse potential (\ref{morse}) (right, between $q_{\rm min}$
and $q_{\rm max}$).}
\label{qinstmoinv}
\end{figure}

If $U_{d}(0)$ is infinite, $T_{\rm c}$ is
formally also infinite; this means that the uniform solution is
unstable at small $q$ for {\it any} temperature (see fig.
\ref{qinstmoinv}, left). For instance, for the
one-dimensional Coulomb potential (\ref{coul1d}), one has (after
regularization) $U_{1}(\kappa)=2/\kappa^{2}$, so that:
\begin{equation}
     q_{\rm max}\,=\,\xi^{-1}\,\sqrt{\frac{2n_{0}\xi v_{0}}{k_{\rm
     B}T}}\enspace,\quad\quad \forall T\enspace.
    \label{qmaxcoul1d}
\end{equation}
The same holds true for higher dimensions, since $U_{d}(\kappa)$ is
for any $d$ proportional to $\kappa^{-2}$.

\vspace{-10pt}
In the preceding examples, the first instability interval, when it
exists, arises
around $q=0$. This is, as explained above, a physical consequence of the
fact that the potential is purely
attractive (the fact that $T_{\rm c}$ is finite or not is a consequence of the
range only). Another interesting
situation is when $v$ is attractive at short distance and repulsive at
long distance; this means that two particles are bound by a potential
barrier but can dissociate when one of them is given enough energy. As
an example, let us consider the {\it inverted} 3$d$ Morse  potential:
\begin{equation}
     v(x)\,=\,v_{0}\,\left[\E^{-2(r-r_{0})/\xi}-2\E^{-(r-r_{0})/\xi}\right]
     \enspace,\hspace{20pt}
     v_{0}\,<\,0
    \enspace.\label{morse}
\end{equation}
In such a case, $T_{\rm c}$ is again finite, and the instability
interval for $q$ is of the form $[q_{\rm min},\,q_{\rm max}]$; it grows
around a finite value (which gives
$U_{3}(q\xi)$ its secondary minimum) and is a consequence, at intermediate
wavelengths, of the interplay between attraction at short distances and
repulsion at large ones. This instability domain is
schematized in figure \ref{qinstmoinv}. Quite naturally, the uniform
state is stable against long wavelength deformations (due to the long
distance repulsive behaviour of the potential) and unstable otherwise.

\section{Numerical study of the Gaussian attractive case}

From the arguments and results given in the two previous sections, we
expect that for a Gaussian attractive (an
example of short-range potential), two
equilibrium solutions exist; the uniform one, which is unstable
against long wavelength deformations when $T\,<\,T_{\rm c}$, where
$T_{\rm c}$ is given by (\ref{Tcgaussd}), and a non-uniform
distribution, which we were unable to find analytically in a closed
form. The aim of this section is to give numerical results for the
one-dimensional gaussian potential. It is shown below that
the predicted transition between these two states is indeed quite sharp
when $T$ crosses $T_{\rm c}$ and that, in addition, an hysteretic
behaviour occurs.

We choose $v({x}) = v_{0}\,\E^{- x^2/(2 \xi^2)}$
($v_{0}<0$) which yields ${\tilde v}({q})=\sqrt{2\pi}\,\xi v_{0}\,
\E^{-\xi^2
{q}^2/2}$. The unstability condition (\ref{instable}) gives the
critical temperature (see (\ref{TC})):
\begin{equation}
     T_{\rm c}\,=\,\sqrt{2\pi}\,n_{0}\,\xi\,|v_{0}|/k_{\rm B}
    \enspace.\label{DCgauss}
\end{equation}

In order to analyze the equilibrium state(s) $n(x)$ arising in the mean-field
treatment with the gaussian potential, we used two numerical
procedures. The first one is a self-consistent iterative procedure  for the
equilibrium equation
(\ref{FPEMFJ0}), hereafter called static algorithm. The second one
(called dynamical algorithm) numerically solves the one-dimensional version
of the full equation (\ref{FPEMFdens}), starting from a given
initial condition $n(x,\,t=0)$, and gives the non-equilibrium
evolution of the system.

With the static algorithm, we start from a
trial density function $n^{(0)}$ and then iterate according to:
\begin{equation}
          n^{(k+1)}(x)=Z^{-1} \E^{-\beta \int_{L} {\D y \,\, n^{(k)}(x)
          v(y-x)}}
\enspace,
\end{equation}
where $Z$ is the normalisation constant. Convergence is reached when
the following quantity:
\begin{equation}
    \int_{L} {\D x \,\, \left(
     \frac{n^{(k+1)}(x)-n^{(k)}(x)}{n^{(k+1)}(x)} \right)^2}
\enspace
\end{equation}
becomes much less than unity. Strict boundary conditions
have been used, namely $j(0,t)=j(l,t)=0$ .

\begin{figure}
\vspace{-5pt} \centering\epsfig{file=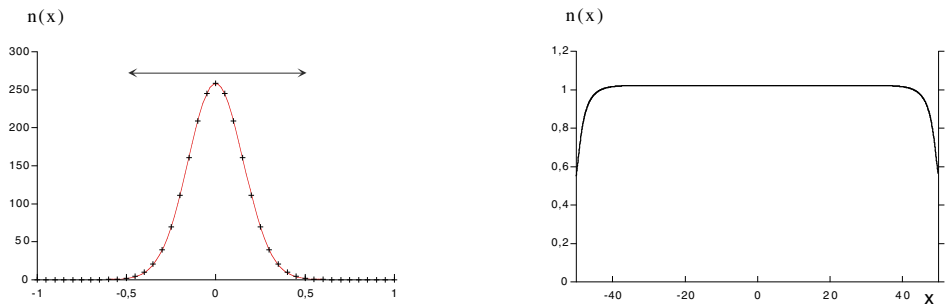,scale=.8}
\vspace{-60pt}
\caption{Equilibrium states for the gaussian potential
(left: $T/T_{c}=0.9$, right: $T/T_{c}=1.1$); the length of
the box is $L=100 \xi$ and $N=100$ ($n_{0}=1$). On the left curve, the crosses
are the results from the static algorithm whereas the solid line is the
gaussian
approximation defined below. $\xi$ is the unit length; note the different
horizontal and vertical scales. Above the localized curve is shown the
lenght $\xi$.}
\label{equilgauss}
\end{figure}

These calculations allow first to check the existence of the
sharp transition occurring at $T_{\rm c}$ when the temperature is
decreased starting with the uniform state; second, they reveal a
hysteretic behaviour: when the temperature is {\it increased} starting from
the localized state (one cluster), the transition from the localized state
to the uniform
one occurs at
another temperature $T^{*}_{\rm c}\,>\,T_{\rm c}$.

Fig. \ref{equilgauss} displays
two typical density profiles (uniform {\it vs} localized cluster),
just above and just below $T_{\rm c}$ obtained with this static
algorithm. Since the width of the localized state turns out to be
a little bit smaller than the range $\xi$ of the potential, the
central part of the cluster can be safely approximated by a gaussian;
indeed, when one expands
the potential near $x/\xi \sim 0$, the harmonic potential is
recovered (if instead of a gaussian potential, one
chooses $v(x)=v_{0} \E^{-|x|/\xi}$, the central shape of the localized
cluster is
$1/\cosh^2 x$, as it must (see (\ref{pcoul1d})) since near the center this
potential is essentially the Coulomb one). Obviously, such an approximation
cannot properly represent the tails of the equilibrium density.

In order to characterize the transition, we introduce two quantities,
$\varepsilon$ and $\lambda$ defined as follows:
\begin{equation}
     \varepsilon(T/T_{\rm c}) = \frac{1}{N} \int_{L} \D x\,\D x'
     \, n(x) n(x') v(x - x')
     \enspace,\label{epsil}
\end{equation}
and:
\begin{equation}
     \Delta^{2}(T/T_{\rm c}) \,= \,\frac{1}{N} \int_{L} \D x \,
     x^2 \,n(x) - \left[\frac{1}{N} \int_{L}\D x\, x\,n(x)\right]^2
     \enspace.
     \label{lbda}
\end{equation}
$\varepsilon$ represents the mean-field interaction energy per particle;
$\Delta$ measures the spatial width of the cluster. For the uniform
state one has $\varepsilon_{\rm unif} \,=\, (N/L) \int_{L} {\D x\,
v(x)}\,\sim\,N(\xi/L)\,v_{0}$
and $\Delta _{\rm unif}\, \sim L \,\gg \xi$; on the contrary, for the
localized state,
$\varepsilon_{\rm loc} \,\sim\,N\,v_{0}$ and $\Delta_{\rm loc}\, \lesssim\,
\xi$. For an infinite system, the ratio
$\varepsilon_{\rm loc}/\varepsilon_{\rm uniform}$ vanishes, so that
$\varepsilon$ can play the role of an order parameter.

The hysteresis is obtained as follows. For each given
temperature, the starting point is the solution found in the previous run
at a slightly
different temperature. After a first
sequence where
$T/T_{\rm c}$ is decreased step by step down to $0.2$, the procedure is
reversed,
$T/T_{\rm c}$ is increased, the iterative procedure starting again from the
localized solution obtained for the previous value of $T/T_{\rm c}$. It is
found that the inverse transition does not occur at $T/T_{\rm c}=1$, but
at a {\it higher} temperature $T_{\rm c}^{*}$.

\begin{figure}
\vspace{-5pt} \centering\epsfig{file=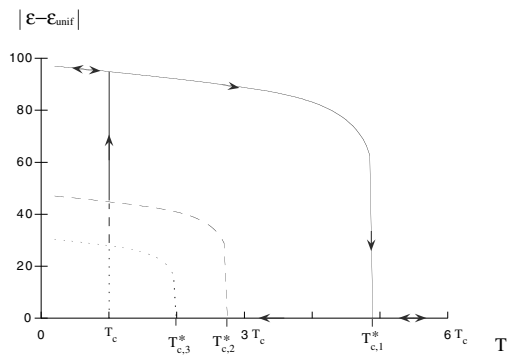,scale=.4}
\caption{$|\varepsilon - \varepsilon_{\rm unif}|$ {\it versus} $T$:
hysteresis
cycle for the uniform-localized transition.  Solid line: 1-peak branch,
dashed line: 2-peak branch, dotted line: 3-peak branch.}
\label{ephist}
\end{figure}

Clearly, for $T<T_{\rm c}$, several solutions can exist. Since the width of
the single peak solution is a bit smaller than the range  $\xi$ of the
potential, one can guess that there also exists a solution with two
peaks separated by a distance much larger that $\xi$. This is
confirmed by the numerical calculations: starting with a trial
function having two separate peaks, the calculation converges towards
a stable solution having the same features. One thus can claim that,
on the low-temperature side, localized solutions exist displaying $j=1,\, 2,\,
3,\,
\ldots$ peaks. Each of these solutions has its own temperature $T_{{\rm
c},\,j}^{*}$. The corresponding typical hysteretic cycles are shown on
figure (\ref{ephist}) for the ``order parameter'' $\varepsilon$ for
the one-,two-, and three-peak solutions.

The possibility of the $n$-peak solutions is confirmed
by the dynamic algorithm (it was checked that the latter eventually
yields the same
equilibrium solutions as the static one). This
allows to conclude that a given $j$-peak solution is linearly
stable for $T_{\rm c}<T<T^{*}_{{\rm c},\,j}$.
Note that the linear stability of the $n$-peak solutions above
$T_{\rm c}$ is not in contradiction with the results obtained in
section \ref{stablan}; all this simply means that, in the region $T_{\rm
c}<T<T^{*}_{{\rm
c},\,j}$, both the uniform state and the localized state(s) are
stable against {\it small} perturbations.
 Defining the entropy as
$S=-\,\int_{L}  \D x\,n(x)\,\ln [n(x)/n_{0}]$, it is seen that the one-peak
solution is the one having the lowest free energy $F=\varepsilon-TS$.
It must be noted that each kind of states has its own relevant physical
parameters. For the uniform state, the relevant
parameters are $v_{0},\xi,n_{0}=N/L$ and the critical temperature found by
the linear
stability analysis is a function of these
parameters. For the localized state, the relevant parameters
are $v_{0},\xi$ and $N$, the number of particles. Indeed, for this
state, edge-effects become irrelevant when $\xi\ll L$: in
this sense $L$ becomes arbitrary so that $n_{0}$ and $N$ are in fact
independent variables.

The physical reason of the hysteresis can be understood as follows.
In the uniform state, a given particle interacts with a small
(microscopic) number
of particles ($\sim n_{0} \xi$); on the contrary, in the localized state, a
single particle interacts with a large number ($\sim N$), macroscopic
in the sense that all the particles effectively interact with anyone
of them (note that this also originates from the fact that the
particles are of zero radius).
The condition for the uniform-to-localized transition is expressed by
$k_{\rm B}T_{\rm c}\,\sim\,n_{0}\xi |v_{0}|$ (see (\ref{TC})). For the
inverse transition, one can expect that a somewhat similar condition
also holds true, so that
$T_{\rm c}^{*}$ is an increasing function of $N$ (and is independent
of $n_{0}$). It is likely that with pointlike
particles $T_{\rm c}^{*}$ goes to infinity with $N$; on the contrary,
with finite-size particles, one can figure out that $T_{\rm c}^{*}$
saturates at very large $N$. As a whole, the hysteretic behaviour is
a consequence of this parameter cross-over between the two kinds of
steady states.

\begin{figure}
\vspace{-5pt} \centering\epsfig{file=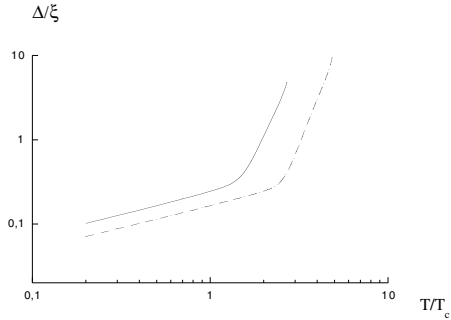,scale=.35}
\caption{ $\Delta$ {\it versus} $T/T_{\rm c}$, for $N=50$ (solid line), and
$N=100$ (dashed line). The two curves display two well
separated regimes; at low temperature, $\Delta \propto T^{1/2}$,
whereas at high temperature, $\Delta$ still follows a
power law, but with an exponent $\simeq 5.4$.
Both curves end at their own localized-uniform transition temperature.}
\label{delta}
\end{figure}

These facts are illustrated by looking at the width $\Delta$ of the
cluster in the localized state, displayed in fig.
\ref{delta} where $\Delta$ is plotted as a function of the relevant variables
for
the uniform state (as they naturally arise in the linear stability
analysis of section \ref{stablan}). Indeed, choosing $N=50$ or $N=100$
with fixed $n_{0}$ gives
two different curves.

\begin{figure}
\vspace{-5pt} \centering\epsfig{file=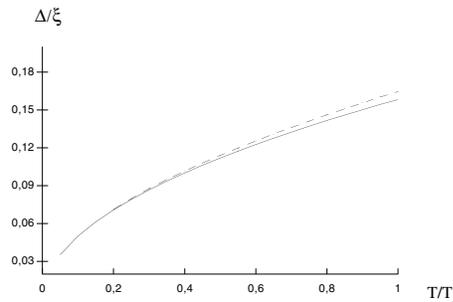,scale=.35}
\caption{Comparison between the gaussian approximation
(dashed line) and the numerical results (solid line) for
the dependence of the width {\it vs} the temperature, in the
low-$T$ region ($N=100$).}
\label{regim}
\end{figure}

In the localized state, the width $\Delta$ of the cluster
displays a cross-over between two regimes. On the low-$T$ side,
one finds approximately $\Delta\propto T^{1/2}$.
This behaviour is the
same as for the harmonic case (see \ref{harmonic}),
which suggests that a gaussian approximation is valid in the low-$T$ region.
Indeed, one can compare $\Delta$ with the analytic expression
(\ref{pharm}) obtained for the harmonic
case (see figure \ref{regim}).
The agreement turns out to be quite good and
this allows to claim that a
gaussian approximation provides a very accurate description of the
localized state arising in the low-$T$ region.
For the second regime ($T_{\rm c}\,<\,T\,<\,T_{\rm c}^{*}$), the
width of the cluster becomes of the same order or even greater
than the potential range, so that the effective potential cannot be well
represented by a harmonic approximation.
Indeed, the increase of the cluster width
with $T$ no more displays a gaussian behaviour, but still
follows a power law.

\section{Dynamics of a system of particles interacting
through the gaussian potential}
\label{Dynamic}

In the previous mean-field analysis, it was shown that
brownian particles interacting through a gaussian potential can have
several steady states when $T_{\rm c}<T<T_{\rm c}^{*}$. The system
will converge towards one of those states, depending of the initial condition.
It can be expected that, under a strong perturbation, the former is able to
shift from one steady
state to another, without changing the two phase space parameters
($T/T_{\rm{c}}$ and $N$).

We here give a few results obtained with the dynamical algorithm for a set of
$N=100$ particles starting in the uniform state at $T/T_{\rm{c}} = 0.5$. The
length of the box is $L=100\xi$. This represents a case where the system,
being in
its stable uniform state at a given $T>T_{\rm c}$, is cooled infinitely
quickly below its critical temperature.

\begin{figure}
\vspace{-5pt} \centering\epsfig{file=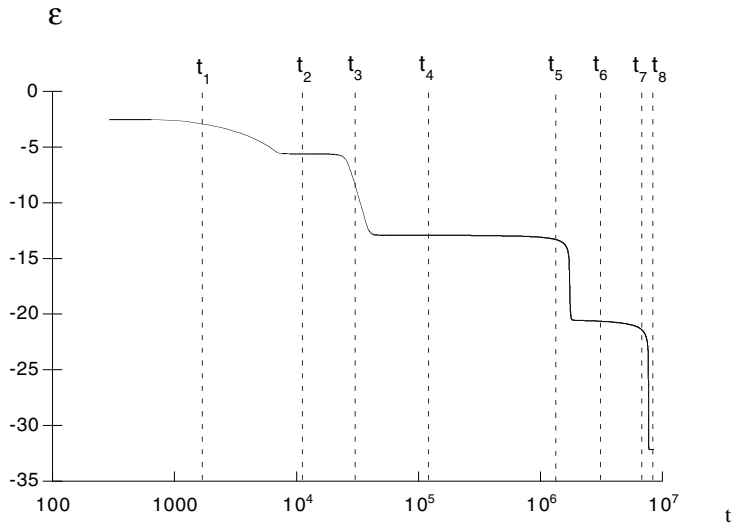,scale=.6}
\caption{Time evolution of the mean energy per particle $\varepsilon$,
starting with the
uniform state sudenly cooled below the critical temperature ($T/T_{\rm c}= 0.5$
and $N=100$). Note the horizontal logarithmic scale; the time unit is
$\xi/(\mu n_{0}|v_{0}|)$.}
\label{fall}
\end{figure}

\begin{figure}
\vspace{-5pt} \centering\epsfig{file=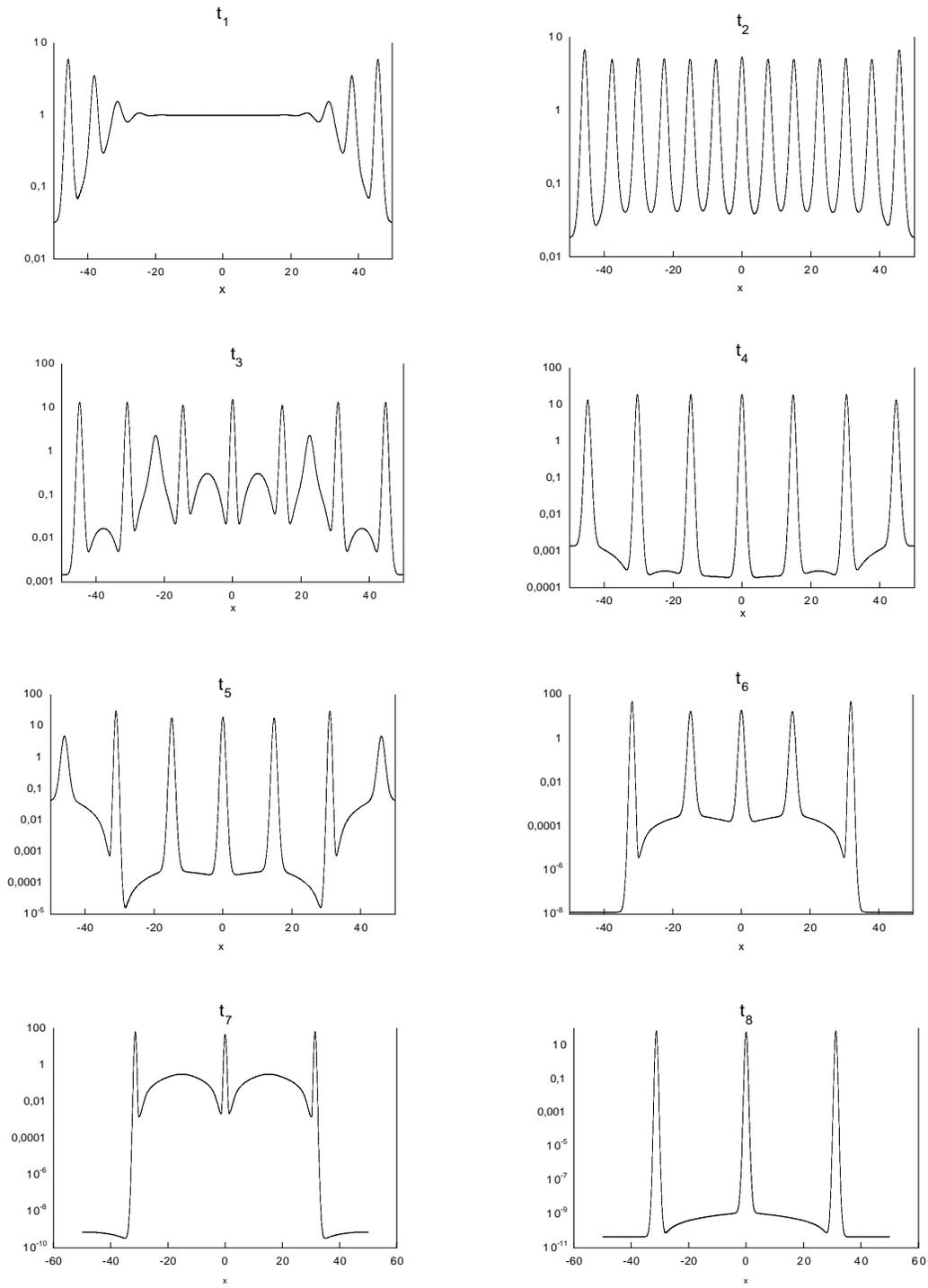,scale=.8}
\caption{Sequence of the density profile at different times, starting from
the uniform state cooled below
$T_{\rm c}$.}
\label{evol}
\end{figure}

We find that the dynamics is a succession of steps, each of them being
a metastable state. Each {\it plateau} has a free energy
lower than that of the preceding. The mean energy per particle is plotted
in fig. \ref{fall}
as a function of time. In order to be complete, some corresponding density
profiles at different
times are shown in fig. \ref{evol}.

These plots show that the dynamics is a succession of metastable
states with a given number of peaks of high density, ($13$-peak state,
$7$-peak state, $5$-peak state,
$3$-peak state). The lifetime of these states rapidly increases when the
number
of peaks gets lower (note the horizontal logarithmic scale in fig. \ref{fall}).

Because of computational time, we were unable to obtain the $1$-peak state.
From static
computations, it is believed that this state is expected to be the final
one, although there is no
evidence  that the two algorithms have the
same attraction basins. Moreover, the $1$-peak state is
not the only stable state: we have checked that also exist stationary
$2$-peak, and
$3$-peak states, a fact which seems nearly obvious on physical grounds
since when the
distance between two neighbouring peaks is much larger that the range $\xi$
no dynamics can occur.

From the time-dependent density profiles (fig. \ref {evol}), we see that
``nucleation''
always occurs at the boundaries of the box. This is easily understood, since a
particle near this boundary is pulled from one side only, as opposed to a
particle located in the bulk. This effect would disappear with periodic
boundary conditions, indicating that the nucleation can only occur around
an inhomogeneity.

It is also interesting to follow the annihilation of the peaks. As an example,
one goes from $13$ to $7$-peaks
{\it via} the diffusion of the even peaks into the odd ones.
Generally speaking, it turns out that the transition towards a state having
fewer
peaks proceeds
through the absorption of ``weak'' peaks into strong ones.
The actual value of the density between the peaks, even when it is
extremely small, turns out to be of first
importance in this mechanism of coalescence.

\section{Conclusion}

In this paper, we have studied properties of the $N$-body
Fokker-Planck equation within a mean field aproximation.
Our aim was to analyse the properties of the steady state -- which is a
uniform density gas when
interactions are set to zero -- as a function of the features of the
interaction
potential. The linear stability analysis of the uniform state allowed to
state the condition
for a phase transition, as a result of the competing effects due to attractive
interaction and to diffusive spreading. In particular, it was shown that
the uniform
phase is always unstable for non-summable attractive potentials, the critical
temperature being infinite. In this latter case, analytical results have
been given for
specific potentials ($1d$-coulombian, harmonic, polynomials), all yielding
an aggregated steady state.

For an attractive short ranged interaction, the above stability condition
was interpreted as a phase transition occurring at a finite $T_{\rm c}$.
This transition
was more thoroughly
studied by numerical computations in a definite case (gaussian potential) and
turns out to be of the first order and hysteretic, the low temperature state
being an inhomogeneous state having one or several peaks of high density.

The far-from-equilibrium dynamics was numerically computed for the gaussian
potential, starting from the
uniform state suddenly cooled below $T_{\rm c}$.
The relaxation proceeds through a sequence of {\it plateaux} displaying a
given
decreasing
number of peaks, each of the former having
a longer and longer lifetime as time goes on. Each of these steps can be
viewed
as a metastable state.

Obviously enough, the considered problem requires further investigations.
One of
them is the relevance of mean field approach. If one expects that
mean field results should be basically valid for long ranged potentials and
high
space dimensionality, their correctness for short range potential has to be
checked by using  more refined
approximations from $N$-body general methods.

\acknowledgements

We are indebted to Julien Vidal for a careful reading of the manuscript and
for
his most valuable remarks.


\end{document}